\PassOptionsToPackage{table,dvipsnames}{xcolor}
\documentclass{article}

\usepackage{arxiv}
\usepackage{algorithm}
\usepackage{algorithmic}
\usepackage{float}
\floatstyle{ruled}
\newfloat{algorithm}{htbp}{loa}
\floatname{algorithm}{Algorithm}
\usepackage[utf8]{inputenc} 
\usepackage[T1]{fontenc}    
\usepackage{hyperref}       
\usepackage{url}            
\usepackage{booktabs}       
\usepackage{amsfonts}       
\usepackage{nicefrac}       
\usepackage{microtype}      
\usepackage{lipsum}		
\usepackage{graphicx}
\usepackage{natbib}
\usepackage{doi}
\usepackage{multirow}
\usepackage{latexsym,amssymb,amsmath,epsfig,bm,psfrag}
\usepackage{floatflt}
\usepackage{lipsum}
\usepackage{xcolor} 

\usepackage{wrapfig}
\usepackage{caption} 
\usepackage[utf8]{inputenc}
\definecolor{bet}{HTML}{BF3131}
\definecolor{wor}{HTML}{48A6A7}
\usepackage{graphicx} 
\usepackage{array}    
\DeclareMathOperator*{\argmax}{arg\,max}
\DeclareMathOperator*{\argmin}{arg\,min}

\usepackage{textcomp}

\definecolor{c1}{HTML}{E7F2E4}
\definecolor{c2}{HTML}{B2C6D5}
\definecolor{bet}{HTML}{FF0B55}
\definecolor{wor}{HTML}{3D90D7}
\definecolor{1st}{HTML}{e6a39d}
\definecolor{2nd}{HTML}{fcf5c1}
\usepackage[table]{xcolor}
\usepackage{colortbl}

\title{LO-SDA: Latent Optimization for Score-based Atmospheric Data Assimilation
}

\author{ {Jing-An Sun$^{1,2,*}$, Hang Fan$^{3,*}$, Ben Fei$^{2,4,\dagger}$, Junchao Gong$^{2,5}$, Kun Chen$^{1,2}$, Fenghua Ling$^{2}$,} \\ \textbf{Wenlong Zhang$^{2}$, Wanghan Xu$^{2}$, Li Yan$^{1}$, Pierre Gentine$^{3}$, Lei Bai$^{2}$}\\
	$^1$ Fudan University, $^2$ Shanghai Artificial Intelligence Laboratory, $^3$ Columbia University, \\ $^4$ The Chinese University of Hong Kong, $^5$ Shanghai Jiaotong University\\
	\texttt{jasun22@m.fudan.edu.cn, benfei@cuhk.edu.hk, baisanshi@gmail.com}}

\date{}

\begin{document}

\maketitle

\begin{abstract}
Data assimilation (DA) plays a pivotal role in numerical weather prediction by systematically integrating sparse observations with model forecasts to estimate optimal atmospheric initial condition for forthcoming forecasts. 
Traditional Bayesian DA methods adopt a Gaussian background prior as a practical compromise for the curse of dimensionality in atmospheric systems, that simplifies the nonlinear nature of atmospheric dynamics and can result in biased estimates.
To address this limitation, we propose a novel generative DA method, LO-SDA. First, a variational autoencoder is trained to learn compact latent representations that disentangle complex atmospheric correlations. Within this latent space, a background-conditioned diffusion model is employed to directly learn the conditional distribution from data, thereby generalizing and removing assumptions in the Gaussian prior in traditional DA methods. Most importantly, we introduce latent optimization during the reverse process of the diffusion model to ensure strict consistency between the generated states and sparse observations. 
Idealized experiments demonstrate that LO-SDA not only outperforms score-based DA methods based on diffusion posterior sampling but also surpasses traditional DA approaches. 
To our knowledge, this is the first time that a diffusion-based DA method demonstrates the potential to outperform traditional approaches on high-dimensional global atmospheric systems. 
These findings suggest that long-standing reliance on Gaussian priors—a foundational assumption in operational atmospheric DA—may no longer be necessary in light of advances in generative modeling.

\end{abstract}

\newcommand\blfootnote[1]{%
\begingroup
\renewcommand\thefootnote{}\footnote{#1}%
\addtocounter{footnote}{-1}%
\endgroup
}
\blfootnote{{$*$}Equal contribution, {$\dagger$}Corresponding author.}

\section{Introduction}
\label{sec:introduction}

\newcommand{\fh}[1]{\textcolor{blue}{#1}}

In numerical weather prediction, data assimilation (DA) is essential for generating accurate initial conditions that directly determine forecast skill~\cite{lorenc1986analysis,gustafsson2018survey,asch2016varda}. Modern DA methods estimate the optimal atmospheric state $ \pmb x $ within a Bayesian framework by combining sparse observations $ \pmb y $ with model forecasts $ \pmb x_b $ (also known as background fields)~\cite{asch2016varda,rabier2003varda,carrassi2018varda,le1986variational}. 
Specifically, DA aims to estimate the Bayesian posterior distribution $ p(\pmb x | \pmb x_b, \pmb y) $. 
Given that forecasts and observations are typically conditionally independent, the posterior simplifies to $ p(\pmb x | \pmb x_b, \pmb y) \propto p(\pmb y | \pmb x) p(\pmb x | \pmb x_b)$.

Traditional DA methods typically assume both the prior $p(\pmb x \mid \pmb x_b)$ and the likelihood $p(\pmb y | \pmb x)$ follow Gaussian distributions to simplify the inference process~\cite{bannister2017review}. 
While this assumption is relatively reasonable for observation errors, it breaks down for background uncertainty, which often becomes non-Gaussian after undergoing nonlinear model evolution.
Furthermore, this assumption makes traditional DA methods rely on the background error covariance matrix $\mathbf{B}$ to define the solution space for assimilation~\cite{bannister2008review1}. Nevertheless, $\mathbf{B}$ often spans more than \( 10^{12} \) degrees of freedom in high-resolution systems, making it extremely challenging to estimate and potentially introducing significant additional error into the assimilation process~\cite{bannister2017review,bannister2008review2}. These limitations have spurred generative DA frameworks.


Generative DA models perform posterior inference using score functions, offering a promising alternative to traditional approaches by relaxing the need for Gaussian assumptions~\cite{rozet2023sda,rozet2023mlpsda,qu2024qusda,huang2024diffda,manshausen2025sda}. However, existing approaches face notable limitations, both in practical implementation and theoretical understanding. 
For instance, Huang et al. (DiffDA)~\cite{huang2024diffda} condition the diffusion model on the background and incorporate observations through a repainting strategy, but their method underperforms in sparse observation settings and cannot effectively handle nonlinear observation operators such as satellite radiative transfer. 
Qu et al.~\cite{qu2024qusda} encode background and multi-modal observations into a unified guidance signal, though their reliance on specific observation distribution assumptions restricts generalization to complex DA scenarios. 
Moreover, Rozet et al.~\cite{rozet2023sda,rozet2023mlpsda} and Manshausen et al.~\cite{manshausen2025sda} treat observations as guidance during the reverse process, ignoring the background prior. While this works well when observations are dense and clean, but often fails under sparse or noisy conditions, where background information becomes essential.
These limitations underscore the necessity of a unified framework that jointly leverages both background information and observational guidance in generative DA.

\begin{figure}
    \centering
    \includegraphics[width=1.0\linewidth]{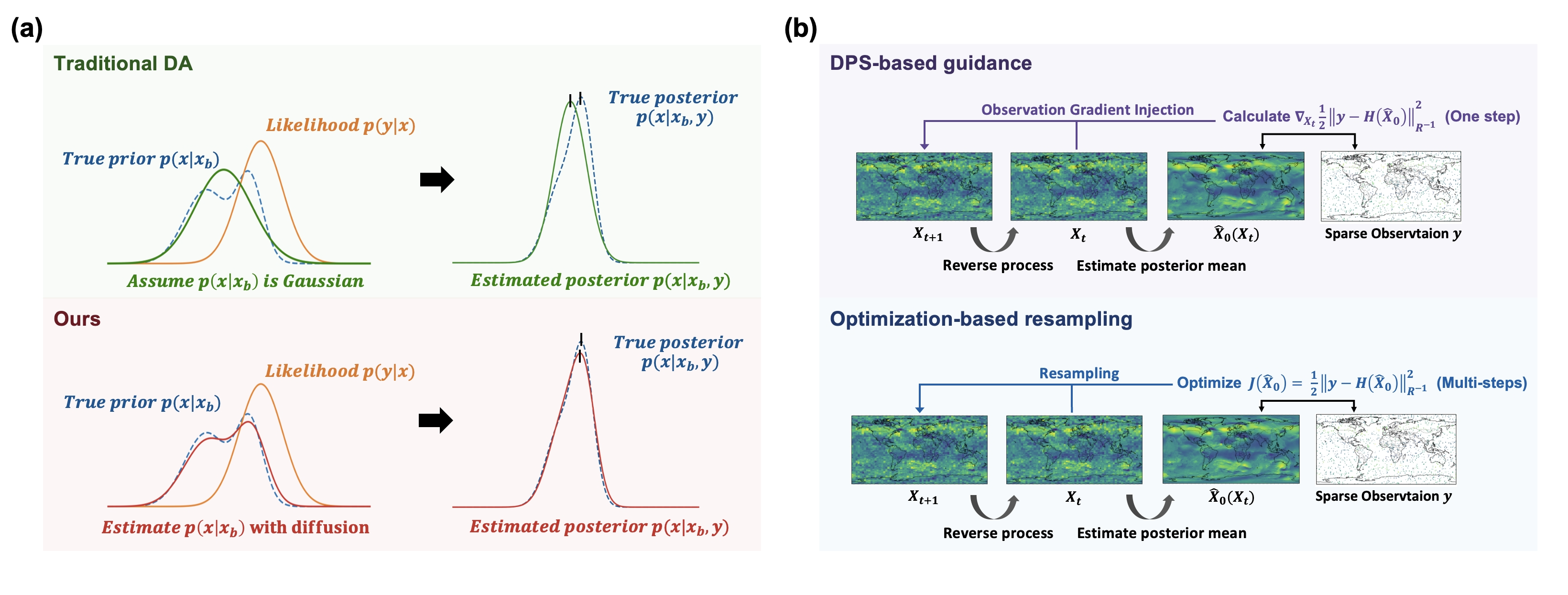}
    \caption{Comparison between LOSDA and other DA approaches. (a) Prior estimation: The true background conditional prior $p(\pmb x|\pmb x_b)$ (blue dashed) is approximated as Gaussian in traditional DA (green), while LOSDA directly estimates it through diffusion modeling (red). By incorporating observation likelihood $p(\pmb y|\pmb x)$, LOSDA achieves posterior estimation $p(\pmb x|\pmb x_b,\pmb y)$ closer to the ground truth. (b) Observation integration methods: Top - Diffusion Posterior Sampling (DPS) updates denoised $\pmb x_t$ via observation error gradient guidance (single-step consistency). Bottom - LOSDA's optimization approach directly minimizes observation error for optimal denoised $\pmb x_t$ (strict multi-step consistency). Our framework enforces tighter observation constraints than gradient-based DPS. }
    \label{fig:overview}
\end{figure}

To this end, we propose the Latent Optimization Score-based Data Assimilation (LO-SDA) framework, which seeks to offer a more principled and reliable formulation of generative DA.
First, we train a variational autoencoder (VAE) to learn a compact latent representation of the high-dimensional atmospheric states, capturing nonlinear dependencies among variables and enabling more efficient probabilistic modeling.
Second, we train a score-based model to model the background conditioned prior in latent space $p(\pmb z|\pmb z_b)$, where $\pmb{z}$ represents the latent representation of model state $\pmb{x}$.
Third, we introduce an alternating latent optimization scheme~\cite{song2024LO} that iteratively enforces observational constraints during guided diffusion sampling. 
In our framework, the diffusion-estimated prior yields a less biased analysis compared to traditional approaches (Figure~\ref{fig:overview} (a)), while latent optimization significantly enhances analysis-observation consistency (Figure~\ref{fig:overview} (b)).
Our framework behaves similarly to multiple posterior likelihood maximization during guidance sampling, distinguishing it from single-step gradient descent in DPS. Meanwhile, our framework has the shared optimization-based strategy employed by variational DA. This mechanism offers a plausible explanation for our framework's superior performance. 

Our contributions are outlined as follows:
\begin{itemize}
    \item We identify a theoretical connection between latent optimization and variational DA methods. Building upon this theoretical analogy, we develop a novel framework that integrates observational information into the background conditional prior through latent optimization techniques. Our alternating latent optimization scheme provably achieves multiple maximizations of the posterior likelihood $p(\boldsymbol y|\boldsymbol z)$ during guided sampling, guaranteeing progressive refinement.
    \item To the best of our knowledge, this is the first work to demonstrate that removing Gaussian assumptions via diffusion enables score-based DA to outperform traditional methods in high-dimensional global atmospheric settings.
    \item By incorporating latent optimization into score-based DA, we iteratively enforce observational constraints during sampling, resulting in more accurate and observation-consistent analyses than those produced by existing approaches.
\end{itemize}

\section{Related work}
\label{sec:related}

\textbf{The variational assimilation methods.} 
Variational assimilation is a representative class of traditional DA methods and is widely used in operational numerical weather prediction systems.
In the three-dimensional case, it seeks to maximize the posterior likelihood~\cite{asch2016varda,rabier2003varda,carrassi2018varda}:
\begin{align}\label{eq:opt}
\pmb{x}_a = \argmax_{\pmb{x}} p(\pmb x|\pmb x_b,\pmb y) = \argmax_{\pmb{x}} p(\pmb x|\pmb x_b)p(\pmb y|\pmb x),
\end{align}
where the assumption of independence between observation errors and background errors is applied.
By assuming that the prior distribution $p(\pmb{x} |\pmb{x}_b)$ and the observation likelihood $p(\pmb{y} |\pmb{x})$ follow Gaussian distributions, three-dimensional variational DA (3DVar) is equivalent to minimizing the following cost function:
\begin{align}\label{eq:3dvar}
    J(\pmb x) = \frac{1}{2} (\pmb x-\pmb x_b)^T \textbf{B} ^{-1} (\pmb x-\pmb x_b) + \frac{1}{2} (\pmb y-\mathcal H (\pmb x))^T \textbf{R}^{-1} (\pmb y-\mathcal H (\pmb x))\ .
\end{align}
where $\mathbf{B}$ and $\mathbf{R}$ denote the covariance matrices of background and observation errors, respectively, and $\mathcal{H}$ is the observation operator that maps model states to observation space.
As noted by Bannister~\cite{bannister2008review1,bannister2017review}, $\mathbf{B}$ plays a central role in variational DA by shaping the feasible solution space and promoting physical consistency in the resulting analysis.
In practice, the high-dimensional $\mathbf{B}$ is commonly simplified via a control variable transformation that approximately diagonalizes it~\cite{descombesGeneralizedBackgroundError2015}. Although this facilitates its inversion, the simplified $\mathbf{B}$ may fail to capture the evolving physical consistency of atmospheric states, leading to suboptimal assimilation outcomes.



\textbf{The latent assimilation methods.} 
Recently, latent data assimilation (LDA)~\cite{cheng2024multilda,melinc20243dlda,peyron2021latentlda,amendola2021datalda,zhengGeneratingUnseenNonlinear,fan2025physicallylda,fanNovelLatentSpace2025,fanNovelLatentSpace2025a} has been proposed to apply traditional DA methods with Gaussian priors in a compact latent space learned via autoencoders.
For example, the latent formulation of the widely used 3DVar—referred to as L3DVar—optimizes the following loss function:
\begin{align}\label{eq:lda}
J(\pmb z) = \frac{1}{2} (\pmb{z}-\pmb{z}_b)^T\textbf{B}_z^{-1}(\pmb{z}-\pmb{z}_b)+\frac{1}{2}(\pmb y-\mathcal H(D(\pmb z))^T \textbf{R}^{-1}(\pmb y-\mathcal H(D(\pmb z))).
\end{align}
where $\pmb{z}$ and $\mathbf{B}_z$ denote the latent state and the background error covariance matrix in the latent space, respectively. Several studies have found that $\mathbf{B}_z$ is inherently near-diagonal, as the latent space effectively captures correlations among atmospheric variables. Consequently, LDA can adopt a diagonal $\mathbf{B}_z$, greatly simplifying its implementation~\cite{melinc20243dlda,zhengGeneratingUnseenNonlinear,fan2025physicallylda}. Fan et al.~\cite{fan2025physicallylda} further showed that performing variational assimilation in latent space can outperform its model-space counterpart. Nevertheless, most latent DA methods remains constrained by the Gaussian prior assumption. 
To overcome this limitation, our work also leverages latent representations of high-dimensional atmospheric states, but replaces the Gaussian prior with a more expressive, data-driven distribution modeled by a latent score-based model.

\section{Method}
\label{sec:method}
\subsection{Preliminary}

\textbf{Score-based model.} 
The score-based model, a promising class of the generative models~\cite{BeatGANdhariwal2021diffusion,BeatGANho2022cascaded}, offering high-quality generation and excellent model convergence~\cite{Conversong2021maximum,Converhuang2021variational,Converkingma2021variational}. It comprises a forward process and a reverse process~\cite{Diffho2020denoising,Diffsohl2015deep,Diffsong2020score}. In the forward process, the original data distribution is transformed into a known prior, by gradually injecting noise. Such a process is governed by a stochastic differential equation (SDE) and a corresponding reverse-time SDE~\cite{Diffsong2020score},
\begin{align}
    d\pmb x &= {\bf f}(\pmb x,t) dt + g(t) d\pmb w\\
    d\pmb{x} &= [{\bf f}(\pmb{x}, t) - g(t)^2\nabla_{\pmb{x}} \log p_t(\pmb{x})]dt + g(t)d\bar{\pmb{w}},\label{eq:denoising}
\end{align}
where the reverse SDE transforms the prior distribution back into the data distribution by gradually removing the noise. Here, $\pmb w$ and $\bar{\pmb{w}}$ both represent the standard Wiener processes (Gaussian white noise), with ${\bf f}(\pmb x,t)$ the drift coefficient and $g(t)$ the diffusion coefficient of $\pmb x(t)$. Accordingly, the perturbation kernel from $\pmb x_0$ to $\pmb x_t$ takes form $p(\pmb x_t|\pmb x)\sim\mathcal{N}(\mu(t),\sigma^2(t)\pmb I)$, where $\mu(t),\sigma^2(t)$ can be determined by the  ${\bf f}(\pmb x,t)$ and $g(t)$. In this work, we take the widely used variance-preserving SDE and the cosine schedule for $\mu(t)$~\cite{rozet2023sda}. In the generative diffusion model, the score function $\nabla_{\pmb{x}} \log p_t(\pmb{x})$ can be estimated by a neural network with parameter $\pmb \theta$ via minimizing the denoising score matching loss $\mathcal L_t \equiv \mathbb{E}_{p(\pmb x_t)} ||\pmb s_{\pmb \theta}(\pmb x,t)-\nabla_{\pmb{x}} \log p_t(\pmb{x}|\pmb x_0)||^2$, which theoretically guarantees $\pmb s_{\pmb \theta}(\pmb x,t)\approx\nabla_{\pmb{x}} \log p_t(\pmb{x})$~\cite{Diffsong2020score}. Once we have a trained $s_{\theta} (\pmb x,t)$, the trajectory from the prior distribution to the real data distribution can be determined following Equation~\ref{eq:denoising}.

\textbf{Score-based data assimilation.} 
Data assimilation under this framework reformulates the Bayesian posterior as a composite scoring process:
\begin{align}
\nabla_{\pmb x_t} \log p(\boldsymbol x_t|\boldsymbol x_b,\boldsymbol y) = {\nabla_{\pmb x_t} \log p(\boldsymbol x|\boldsymbol x_b)} + {\nabla_{\pmb x_\tau} \log p(\boldsymbol y|\boldsymbol x_t)}
\end{align}
The prior term leverages the diffusion model's capacity to capture complex spatial correlations, bypassing the oversimplified Gaussian assumptions in the conventional DA methods. The constraint term enforces observation consistency through conditional guidance. In the DPS paradigm~\cite{chung2023diffusionDPS}, the observation term is supposed to follow Gaussian distribution,
\begin{align}
p(\boldsymbol y|\boldsymbol x_t) \sim \mathcal{N}\left(\mathcal{H}(\tilde{\pmb x}_0(\pmb x_t)), \pmb{R}\right)
\end{align}
where the posterior mean $\tilde{\pmb x}_0$ derives from Tweedie’s formula~\cite{Diffho2020denoising,Diffsohl2015deep}:
\begin{align}\label{eq:dps}
\tilde{\pmb x}_0(\pmb x_t) = \frac{\pmb x_t + \sigma(t)^2 \pmb s_{\pmb \theta} (\pmb x_t,\pmb x_b)}{\mu(t)}.
\end{align}

\begin{algorithm}[t]
\caption{Comparison of DPS Guidance and Latent Optimization for Score-Based Data Assimilation}
\label{algo:laop-dps}
\begin{algorithmic}[1]
\STATE \textbf{Input:} Pretrained score function $\boldsymbol{s}_{\boldsymbol{\theta}}(\boldsymbol{z}_t,\boldsymbol z_b) = \nabla_{\boldsymbol{z}_t}\log p(\boldsymbol z_t|\boldsymbol z_b)$, pretrained VAE (encoder $E(\cdot)$, decoder $D(\cdot)$), observation distribution $p(\boldsymbol y|\boldsymbol z_t)$, observations $\boldsymbol{y}$.

\FOR{$t = 1$ to $0$} 
    \STATE Solve reverse SDE with $\pmb z_{t+1}$ and score function $\nabla_{\boldsymbol{z}_t}p(\boldsymbol z_t|\boldsymbol z_b)$: $\tilde{\boldsymbol{z}}_t \gets \text{SolutionAtTime}(t)$
    \IF{$t \in C$} 
        \STATE \text{Calculate posterior mean:} \mbox{$\textstyle\hat{\pmb z}_0\!=\!\frac{\tilde{\boldsymbol{z}}_t + \sigma^2(t) \nabla_{\pmb z_t} \log p(\pmb z_t|\pmb z_b)}{\mu(t)} $}\\
        \colorbox{c1}{
            \parbox{\dimexpr\linewidth-2\fboxsep}{
                \begin{minipage}{\linewidth}
                \STATE \textbf{DPS guidance}
                    \STATE Perform diffusion posterior sampling:
                    \begin{align}
                        \boldsymbol{z}_t &= \tilde{\boldsymbol{z}}_t +\zeta \nabla_{\boldsymbol{z}_t} \log p(\boldsymbol{y}|\boldsymbol{z}_t)\nonumber\\
                        &=\tilde{\boldsymbol{z}}_t -\frac{1}{2}\zeta \nabla_{\tilde{\boldsymbol{z}}_t}(\pmb y-\mathcal H(D(\hat{\pmb z}_0))^T\pmb R^{-1}(\pmb y-\mathcal H(D(\hat{\pmb z}_0))
                        \end{align}
                \end{minipage}
            }
        }

        \colorbox{c2}{
            \parbox{\dimexpr\linewidth-2\fboxsep}{
                \begin{minipage}{\linewidth}
                \STATE \textbf{Latent Optimization} 
                \STATE 
                With initial value $\boldsymbol{z}^0=\hat{\boldsymbol{z}}_0$ 
                    \STATE \textbf{Repeat} 
                    \begin{align}
                        \boldsymbol{z}^{i+1} &= \boldsymbol{z}^{i} +\zeta \nabla_{{\boldsymbol{z}}^i} \log p(\boldsymbol{y}|{\boldsymbol{z}}^i)\nonumber\\
                        &=\boldsymbol{z}^{i} - \frac{1}{2}\zeta \nabla_{{\boldsymbol{z}}^i}(\pmb y-\mathcal H(D({\pmb z}^i))^T\pmb R^{-1}(\pmb y-\mathcal H(D({\pmb z^i}))
                        \end{align}
                    \STATE \textbf{Until} Convergence to $\hat{\boldsymbol{z}}_0(\boldsymbol{y})$
                    \STATE Go back to the noising manifold by resampling: $\pmb z_t \sim p({\boldsymbol{z}}_t | \tilde {\boldsymbol{z}}_t, \hat{\boldsymbol{z}}_0(\boldsymbol{y}), \boldsymbol{y}) $
                \end{minipage}
            }
        }

    \ELSE
        \STATE $\pmb z_t = \tilde{\pmb z}_t$
    \ENDIF
\ENDFOR
\STATE \textbf{Return:} The decoded optimized latent variables $D(\boldsymbol{z}_0)$
\end{algorithmic}
\end{algorithm}


\subsection{Latent score-based data assimilation} 
Due to the computational challenges in high-dimensional systems, LDA~\cite{cheng2024multilda,melinc20243dlda,peyron2021latentlda,amendola2021datalda,fan2025physicallylda} is proposed to leverage VAEs for compressing physical fields into low-dimensional manifolds~\cite{kingma2013autovae,doersch2016tutorialvae} and performing efficient optimization in this reduced space. 
Specifically, the traditional 3DVar formulation (Equation~\ref{eq:opt}) is adapted to the latent space with cost function described by Equation~\ref{eq:lda}. 
The gradient descent iteratively optimizes the latent representation of the analysis field. 
The optimized latent is then decoded to reconstruct the assimilated state. 
Although LDA alleviates non-linearity challenges, its retention of Gaussian assumptions for latent background distributions imposes theoretical limitations as above-discussed, particularly in capturing the multiscale complexity characteristic of real atmospheric states~\cite{fan2025physicallylda}. 
In this work, we train a score-based model in the latent space to model the background conditional distribution. 
Additionally, we integrate observations through guidance sampling in the latent space. Mathematically, the latent score modeling for DA can be expressed as:
\begin{align}
    \nabla _{\pmb z_t} \log p(\boldsymbol z_t|\boldsymbol z_b,\boldsymbol y) &= \nabla _{\pmb z_t} \log p(\boldsymbol z|\boldsymbol z_b)+\nabla _{\pmb z_t} \log p(\boldsymbol y|\boldsymbol z_t)\nonumber\\
    &=\pmb s_{\pmb \theta} (\pmb z_t,\pmb z_b) + \nabla _{\pmb z_t} \log p(\boldsymbol y|\boldsymbol z_t).
\end{align}
For the guidance term, we implemented the latent counterpart of DPS guidance where the observation term preserves Gaussian distributions $p(\boldsymbol y|\boldsymbol z_t)\sim \mathcal{N}(\mathcal H(D(\hat{\pmb z}_0(\pmb z_t)),\pmb R)$. 
Similar to Equation~\ref{eq:dps}, $\hat{\pmb z}_0$ is the posterior mean. $D(\cdot)$ denotes the decoder of VAE. While Algorithm~\ref {algo:laop-dps} outlines the sampling process using DPS guidance, its single-step gradient update mechanism may provide insufficient constraint enforcement, potentially compromising observation consistency in high-dimensional scenarios. 

\subsection{Latent optimization techniques}\label{sec:laop}
To perform strict observation consistency, we aim to integrate variational optimization (Equation~\ref{eq:3dvar} and Equation~\ref{eq:lda}) used in traditional DA. Inspired by inverse problem solving techniques~\cite{song2024LO} within diffusion models, a two-stage latent optimization strategy is proposed: Hard-Constrained Optimization: (1) Solving $\hat {\pmb z}_0(\pmb y)=\argmin_{\boldsymbol z} (\boldsymbol y - \mathcal{H}(D(\boldsymbol z)))^T{\pmb R^{-1}}(\boldsymbol y - \mathcal{H}(D(\boldsymbol z)))$ to ensure strict observation consistency, (2) Projecting the optimized latent back to the noisy data manifold using the reverse process. Since the $\hat {\pmb z}_0(\pmb y)=(\pmb z_t-\sigma(t)\varepsilon)/\mu(t)$ can be viewed as the estimated mean of latent $\pmb z_t$ with the observation single $\pmb y$, one have that 
\begin{align}
    p(\pmb z_t|\hat {\pmb z}_0(\pmb y),\pmb y) \sim\mathcal{N}\left(\mu(t)\hat{\pmb z}_0(\pmb y),\sigma^2(t)\pmb I\right),
\end{align}
from the forward process. 
When we map the $\hat {\pmb z}_0(\pmb y)$ back to noise data manifold, we need the distributions $p({\boldsymbol{z}}_t | \tilde {\boldsymbol{z}}_t, \hat{\boldsymbol{z}}_0(\boldsymbol{y}), \boldsymbol{y})$. By Bayesian formula, 
$p({\boldsymbol{z}}_t | \tilde {\boldsymbol{z}}_t, \hat{\boldsymbol{z}}_0(\boldsymbol{y}), \boldsymbol{y})\propto p(\tilde {\boldsymbol{z}}_t | {\boldsymbol{z}}_t, \hat{\boldsymbol{z}}_0(\boldsymbol{y}), \boldsymbol{y}) p(\pmb z_t|\hat{\pmb z}_0(\pmb y),\pmb y)$. The posterior distribution $p(\tilde {\boldsymbol{z}}_t | {\boldsymbol{z}}_t, \hat{\boldsymbol{z}}_0(\boldsymbol{y}), \boldsymbol{y})$ is assumed as Gaussian distribution with variance $\lambda_t^2$ and the $p(\pmb z_t|\hat {\pmb z}_0(\pmb y),\pmb y)$ is supposed to provide the prior of its mean. 
Thus, it is accordingly derived (see Appendix):
\begin{align}\label{eq:res}
    p({\boldsymbol{z}}_t | \tilde{\boldsymbol{z}}_t, \hat{\boldsymbol{z}}_0(\boldsymbol{y}), \boldsymbol{y}) = \mathcal{N} \left( \frac{\lambda_t^2 \mu(t) \hat{\boldsymbol{z}}_0(\boldsymbol{y}) + \sigma^2(t) \tilde{\boldsymbol{z}}_t}{\lambda_t^2 + \sigma^2(t)}, \frac{\lambda_t^2 \sigma^2(t)}{\lambda_t^2 + \sigma^2(t)} \boldsymbol{I} \right).
\end{align}
Following~\cite{song2024LO}, we choose the variance $\lambda_t^2$ schedule as $\lambda_t^2 = \lambda \left(\frac{1-\mu^2(t-\Delta t)}{\mu^2(t)}\right) \left( 1- \frac{\mu^2(t)}{\mu^2(t-\Delta t)}\right)$ with a hyperparameter $\lambda$. This approach integrates variational optimization within the diffusion sampling framework (Algorithm~\ref {algo:laop-dps}), where latent variables are iteratively refined at multiple diffusion steps, preserving highly observation consistency. 

We identify a potential theoretical connection between the latent optimization technique and the variational DA methods. In variational DA, the analysis field is obtained by minimizing a cost function (Equation~\ref{eq:lda}) that balances the background term and observation posterior likelihood. Analogously, as demonstrated in Algorithm~\ref{algo:laop-dps}, each latent optimization step during reverse sampling updates the latent state through strict observation consistency (corresponding to the observation term in Equation~\ref{eq:lda}). The optimized latent is subsequently fed into the background conditional diffusion kernel
$p(\pmb z_t|\pmb z_{t+1})$ to correct the diffusion trajectory. 
This optimization-based similarity not only motivates us to apply latent optimization to integrate observational information into the background conditional prior but also provides a plausible explanation of our framework's outstanding performance.

\begin{figure*}[t]
    \centering
    \includegraphics[width=1.0\linewidth]{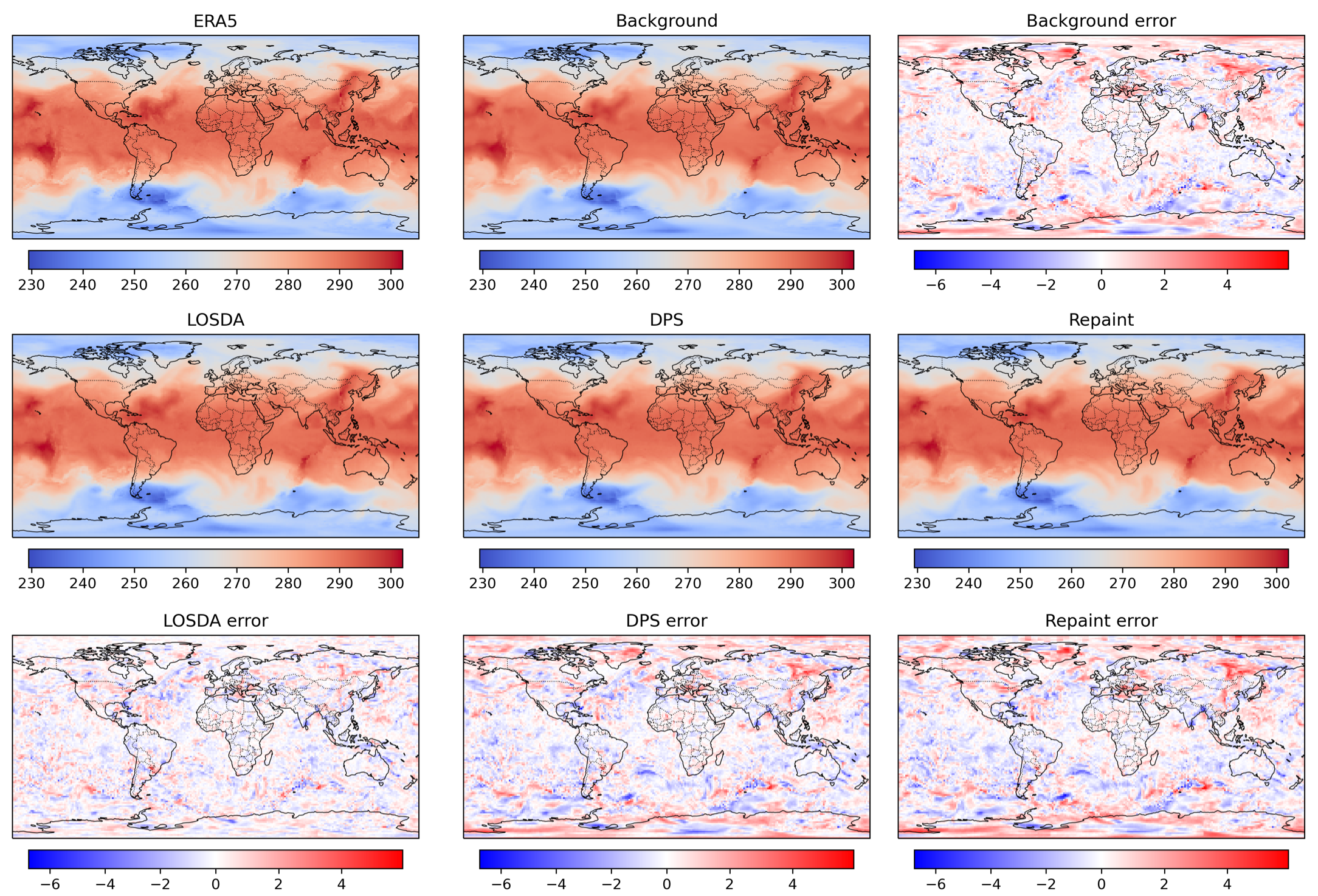}
    \caption{Comparative visualization of t850 analysis fields across assimilation methods under 1\% idealized observation (valid at 2019-01-03 00:00 UTC).
Top row (left to right): ERA5 ground truth, background field, and background error.
Middle row: Assimilation results from (a) proposed LO-SDA method, (b) DPS framework, and (c) Repaint approach.
Bottom row: Corresponding absolute error fields relative to ERA5 truth.
The reduced error magnitude (lighter hues) in LO-SDA results demonstrates our method's superior error reduction capability compared to alternative approaches.}
    \label{fig:enter-label}
\end{figure*}
\section{Experiments}
\label{sec:experiments}

\subsection{Experimental Settings and Evaluations }
\textbf{Dataset and metrics.} We conduct our experiments on the ERA5 reanalysis dataset~\cite{hersbach2020era5}, a global atmospheric data product maintained by the European Centre for Medium-Range Weather Forecasts (ECMWF). Our study utilizes 5 upper-air atmospheric variables (geopotential, temperature, specific humidity, zonal wind, and meridional wind) across 13 pressure levels (50hPa, 100hPa, 150hPa, 200hPa, 250hPa, 300hPa, 400hPa, 500hPa, 600hPa, 700hPa, 850hPa, 925hPa, and 1000hPa), combined with 4 surface variables (10-meter zonal component of wind (u10), 10-meter meridional component of wind (v10), 2-meter temperature (msl) and mean sea level pressure (msl)), forming a total of 69 meteorological variables. The pressure-level variables follow the standardized ERA5 naming convention (e.g., t850 denotes temperature at 850 hPa). We use a subset spanning 1979-2018 for training and evaluations. 
For evaluations, the assimilation quality is assessed by direct comparison with ERA5 reference fields. Three metrics quantifying performance are overall mean square error (MSE), mean absolute error (MAE), and the latitude-weighted root mean square error (WRMSE) (see Appendix), which is a statistical metric widely used in geospatial analysis and atmospheric science~\cite{rasp2020weatherbench,rasp2024weatherbench2}. The validation procedure conducts assimilation cycles at 00:00 UTC for each day throughout 2019. For each test case, we calculate the above three metrics. Final performance scores represent the annual average of these daily metrics, ensuring statistically significant results across all seasons and synoptic conditions.

\textbf{Experimental setting.} The Fengwu AI forecasting model~\cite{chen2023fengwuAIfore} (6-hour temporal resolution) is integrated into our DA framework to produce the background field. These fields are generated through an 8-step autoregressive forecasting procedure, initialized with ERA5 conditions from 48 hours prior to the target assimilation lead time. To simulate realistic observing system characteristics, we create synthetic observations by randomly masking the ERA5 truth data at two sparsity levels (95\% and 99\%), mimicking typical satellite coverage constraints.
The 1.40625° ($128\times256$ grid) spatial resolution is employed, yielding input arrays of size $69\times128\times256$.  





\begin{table*}[t]
\centering
\caption{Quantitative performance comparison of different methods under 1\% and 5\% observations. 
}
\resizebox{\textwidth}{!}
{
\begin{tabular}{c|c|cc|cccccc}
\toprule
\multirow{3}{*}{Ratio}& \multirow{2}{*}{Model} & \multirow{2}{*}{MSE} &\multirow{2}{*}{MAE} &\multicolumn{6}{c}{WRMSE} \\
 & &  &  & msl & u10 & u700 & v500 & z500 & t850  \\\cmidrule{2-10} 
 & 48h background & 0.0505 &	0.1178 & 98.7265&1.2727 &	1.9953&2.4217 &	89.2752 &	0.9310 	\\ \midrule
\multirow{5}{*}{1\% observation} & 3DVAR & 0.0483& 0.1138& 81.6384& 1.2235& 1.9850& 2.4298& 62.9377& \colorbox{1st}{0.8797}  \\ 
& L3DVAR & \colorbox{2nd}{0.0474}& \colorbox{2nd}{0.1105}&  \colorbox{1st}{62.1054} & \colorbox{2nd}{1.1862} & \colorbox{2nd}{1.9797} & \colorbox{2nd}{2.3392} & \colorbox{1st}{53.0902} & \colorbox{2nd}{0.8975} \\ 
& Repaint & 0.0592 &0.1311 &114.3672 & 1.4167&2.1059 & 2.5664& 104.1351& 1.0363  \\
 & DPS &0.0545& 0.1247&95.9850 &1.3286&2.0220 &2.3981& 85.5673& 1.0269  \\ 
 & LO-SDA(ours) & \colorbox{1st}{0.0472} & \colorbox{1st}{0.1101}&\colorbox{2nd}{62.4505} & \colorbox{1st}{1.1836} & \colorbox{1st}{1.8981}&\colorbox{1st}{2.2439} & \colorbox{2nd}{53.6468} & 0.9243  \\  \midrule
 \multirow{5}{*}{5\% observation} & 3DVAR & 0.0430& 0.0982& 63.2562& 1.1661& 1.8765 & 2.2365& 45.8574 & 0.7849 \\ 
 & L3DVAR & \colorbox{2nd}{0.0315}& \colorbox{2nd}{0.0903}& \colorbox{2nd}{45.8037} & \colorbox{2nd}{0.9350} & \colorbox{2nd}{1.7166} & \colorbox{2nd}{1.9400} & \colorbox{2nd}{38.6411} & \colorbox{1st}{0.8024}  \\ 
& Repaint & 0.0496& 0.1219 &106.3938 &1.2934 & 1.9889&2.3622 &  95.9167 & 0.9856 \\
 & DPS & 0.0486 & 0.1199 & 93.247&1.2673 & 1.9585&2.3271 & 85.3329 &0.9891 \\ 
 & LO-SDA(ours) & \colorbox{1st}{0.0309} & \colorbox{1st}{0.0851} & \colorbox{1st}{42.3498} &\colorbox{1st}{0.8992} & \colorbox{1st}{1.5873}& \colorbox{1st}{1.7894} & \colorbox{1st}{32.8990} & \colorbox{2nd}{0.8094}   \\  \bottomrule
\end{tabular}
}
\label{tab:std0_main}
\end{table*}

\textbf{The background conditional diffusion model.}
We present a unified framework for conditional physics field modeling through variational autoencoding and latent diffusion. Our architecture begins with a window-attention transformer VAE~\cite{han2024cra5} that compresses high-dimensional fields ($69\times128\times256$) to compact latent representations ($69\times32\times64$). Trained for 80 epochs using AdamW~\cite{loshchilov2017fixingadam} with batch size 32, the VAE employs a hybrid learning rate schedule: linear warmup to $2\times10^{-4}$ over 10,000 iterations followed by cosine decay, achieving 0.0067 reconstruction MSE as detailed in Appendix. The latent diffusion process then learns conditional distributions $p(\pmb z|\pmb z_b)$ through a 28-layer transformer backbone~\cite{peebles2023scalabledit} with 1152-dimensional hidden states, (2,2) patch embedding, and 16-head cross-attention for background latent $\pmb{z}_b$ conditioning. For diffusion setting, the variance-preserving SDE~\cite{Diffsong2020score} with cosine noise scheduling. Optimized via AdamW~\cite{loshchilov2017fixingadam} at constant $1\times10^{-4}$ learning rate (batch size 32), the model converges stably over 100k training steps. Sampling employs a modified Predictor-Corrector scheme combining 128-step prediction with 2 iterations of Langevin correction~\cite{Diffsong2020score}. See the Appendix for more comprehensive resource usage.

\textbf{Baselines.} For diffusion-based experiments, we incorporate observations through two baseline methods: a latent-space implementation of the repainting technique from DiffDA~\cite{huang2024diffda} and the latent version of DPS described in Algorithm~\ref{algo:laop-dps}. The latent repaint implementation follows:
\begin{align}
    \pmb z^{obs}_{t} \sim \mathcal N(\mu(t)&E(\pmb x^*),\sigma^2(t)\pmb I),\quad \tilde{\pmb z}^t \gets \text{SolutionAtTime}(t)\\
    \pmb z_{t-1} &= E\left(m\odot D(\pmb z^{obs}_{t}) + (1-m)\odot D(\tilde{\pmb z}^t)\right)
\end{align}
where $z^{obs}_{t}$ is noised latent with $
E(\pmb x^*)$ is the encoded ERA5 ground truth latent. $\pmb z_t$ is the sampled prior latent at diffusion time $t$. We finally combine the decoded observed and prior latents in model space using a masking matrix $m$, with $\odot$ denoting element-wise multiplication.

For the DPS assimilation, we apply guidance during the final third of the reverse sampling process with a scale factor of 800. Our LO-SDA implementation employs a hyperparameter setting of $\lambda=100$ to balance observation constraints and prior knowledge. In non-diffusion experiments, we implement both conventional 3DVar and its latent-space counterpart (L3DVar) via optimizing cost function in Equation~\ref{eq:3dvar} and Equation~\ref{eq:lda}, respectively. This dual approach allows us to systematically evaluate the performance improvements offered by latent-space assimilation techniques across different methodological frameworks.

\subsection{Results and Discussions}

Table~\ref{tab:std0_main} provides a quantitative evaluation of analysis errors by comparing LO-SDA with the baseline methods. 
The error of the background field is presented in the first row. 
Among diffusion-based approaches (Repaint, DPS, and LO-SDA), LO-SDA with latent optimization demonstrates superior performance over both repainting and DPS guidance across comprehensive metrics (MSE, MAE, and WRMSE), particularly for most prognostic variables. 
For example, LO-SDA achieves 13.39\% and 20.27\% improvement over DPS and Repaint in overall MSE metrics, aligning with our theoretical expectation in Algorithm~\ref{algo:laop-dps}. 
These results confirm that latent optimization enforces rigorous data consistency.
When compared to traditional 3DVAR data assimilation, LO-SDA exhibits significantly improved accuracy. 
Notably, LO-SDA performs comparably to the ML-enhanced variational method, L3DVAR.
Moreover, LO-SDA demonstrates superior scalability with increased observational density - at 5\% observation coverage, it surpasses L3DVAR's performance, highlighting the effectiveness of latent optimization in leveraging observational constraints.

To further elucidate these findings,
we quantify the percentage improvement relative to 3DVAR and L3DVAR. Under 1\% observation conditions, LO-SDA achieves a 14.76\% improvement over 3DVAR in z500 WRMSE, while maintaining comparable performance ($-1.05\%$) with L3DVAR. This performance gap substantially widens at 5\% observation density, where LO-SDA delivers 28.25\% and 14.86\% improvements over 3DVAR and L3DVAR, respectively, for z500 WRMSE. 
These results underscore two critical findings: (1) The latent optimization framework effectively assimilates observational information, with its advantage becoming more pronounced as observational density increases; (2) LO-SDA exhibits strong scalability, suggesting its suitability for operational implementation with realistic observational networks. 

\begin{table*}[t]
\centering
\caption{Quantitative performance comparison under a 1\% observation setting, with varying observation errors modeled as Gaussian noise. The standard deviations are set to 0.02, 0.05, and 0.10 relative to the ERA5 climatological standard deviation.
}
{
\begin{tabular}{c|c|cc|cccccc}
\toprule
 \multirow{3}{*}{Ratio}& \multirow{2}{*}{Model}& \multirow{2}{*}{MSE} &\multirow{2}{*}{MAE} &\multicolumn{6}{c}{WRMSE} \\
 & &  &  & msl & u10 & u700 & v500 & z500 & t850 \\ \cmidrule{2-10}
 & 48h background & 0.0505 & 0.1178 & 98.7265&1.2727 &1.9953 &2.4217 & 89.2752 & 0.9310  \\ \midrule
 \multirow{3}{*}{std = 0.02 } & 3DVAR & 0.0484& 0.1141& 82.4252 & 1.2243 & \colorbox{1st}{1.9805} & 2.4202 & 67.3522 & \colorbox{1st}{0.8679} \\ 
 & L3DVAR & \colorbox{2nd}{0.0475}& \colorbox{2nd}{0.1109}& \colorbox{1st}{63.0360} & \colorbox{1st}{1.1935} & \colorbox{2nd}{1.9900} & \colorbox{2nd}{2.3496} & \colorbox{1st}{53.6535} & \colorbox{2nd}{0.9054} \\ 
 & LO-SDA(ours) & \colorbox{1st}{0.0470} & \colorbox{1st}{0.1109} &\colorbox{2nd}{64.7142} & \colorbox{2nd}{1.1940} & 2.0618&\colorbox{1st}{2.2888} & \colorbox{2nd}{55.3858} & 0.9354   \\ \midrule
\multirow{3}{*}{std = 0.05} & 3DVAR & 0.0485& 0.1158& 85.4325 & 1.2246 & 1.9643 & 2.3893 & 77.8867 & \colorbox{1st}{0.8845}  \\ 
& L3DVAR & \colorbox{2nd}{0.0481}& \colorbox{2nd}{0.1132}& \colorbox{2nd}{72.2672} & \colorbox{1st}{1.1852} & \colorbox{2nd}{1.9515} & \colorbox{2nd}{2.3199} & \colorbox{2nd}{63.6825} & \colorbox{2nd}{0.9143} \\  
 & LO-SDA(ours) & \colorbox{1st}{0.0479} & \colorbox{1st}{0.1127}&\colorbox{1st}{68.0784} &\colorbox{2nd}{1.1989} &\colorbox{1st}{1.9104} & \colorbox{1st}{2.2641} & \colorbox{1st}{58.1329} & 0.9542   \\  \midrule
 \multirow{3}{*}{std = 0.10 } & 3DVAR & \colorbox{2nd}{0.0495}& \colorbox{2nd}{0.1173}& 91.2331 & \colorbox{2nd}{1.2291} & \colorbox{2nd}{1.9493} & 2.3636 & 84.4765 & \colorbox{1st}{0.9101} \\ 
 & L3DVAR & \colorbox{1st}{0.0489}& \colorbox{1st}{0.1147}& \colorbox{2nd}{83.1325} & \colorbox{1st}{1.1954} & \colorbox{1st}{1.9448} & \colorbox{2nd}{2.3275} & \colorbox{2nd}{75.3260} & \colorbox{2nd}{0.9341}  \\ 
 & LO-SDA(ours) & 0.0498 & 0.1188 &\colorbox{1st}{82.0772} &1.2408&1.9602 & \colorbox{1st}{2.3194} & \colorbox{1st}{69.4214} & 1.0308  \\ \bottomrule
\end{tabular}
}
\label{tab:std_abla}
\end{table*}

The success of LO-SDA demonstrates that generative modeling can enable atmospheric DA transcend traditional Gaussian assumptions.
By learning non-parametric conditional distributions $p(\pmb{x}|\pmb{x}_b)$, our framework achieves more accurate analysis fields than both variational methods and recent diffusion-based DA approaches. 
This advancement stems from two key innovations: (1) a background-conditioned diffusion model that replaces restrictive Gaussian priors, and (2) latent optimization during the reverse process to enforce hard observation constraints - a novel hybrid approach that marries the flexibility of generative modeling with the observation consistency requirements of operational DA. 

\subsection{Real-world observations}
To evaluate our framework under real-world conditions, we employ the Global Data Assimilation System (GDAS) prepbufr dataset, which incorporates multi-source observations. 
For this study, only surface and radiosonde observations are utilized. These observations are first interpolated onto the model state grid, and any multiple observations at a single grid point are averaged. High-elevation surface observations are vertically interpolated and reclassified as upper-air data.
A quality control procedure is further applied to remove observations with large deviations. Observations are dropped if their deviation from the ERA5 reference exceeds 0.05 of the ERA5 climatological standard deviation. We perform data assimilation daily at 00:00 UTC throughout 2017, using a 48-hour background field. As shown in Table~\ref{tab:real}, the results indicate that LO-SDA achieves performance comparable to L3DVAR, and slightly outperforms the traditional 3DVAR when using real observations.

\begin{table*}[t]
\centering
\caption{Assimilation performance on real-world observation across various methods.}
{
\begin{tabular}{c|c|c|cccccc}
\toprule
 & \multirow{2}{*}{MSE} & \multirow{2}{*}{MAE} & \multicolumn{6}{c}{WRMSE} \\
 & & & msl & u10 & u700 & v500 & z500 & t850 \\ 
\midrule
48h background & 0.0475 & 0.1158 & 98.7910  &  1.2588   &1.9692 &  2.4061 & 89.2800  &   0.9232 \\ 
\midrule
3DVAR & 0.0472&0.1150 & 87.1536 & 1.2532 & 1.9646 & 2.3950 & 83.8241 & 0.9068 \\
L3DVAR &0.0467 &0.1143 & 84.8751 & 1.2376 & 1.9597 & 2.3778 & 78.1842 & 0.8962 \\
LO-SDA(ours) & 0.0469 & 0.1140 & 81.4013 & 1.2128 & 1.9891 &  2.3657 & 77.1881 & 0.9914 \\
\bottomrule
\end{tabular}
}
\label{tab:real}
\end{table*}

\subsection{Ablation studies}
\textbf{Observation Error Robustness.} 
To evaluate the robustness of LO-SDA under realistic observational conditions, we conduct experiments with simulated observation errors by injecting additive Gaussian noise with standard deviations of 2\%, 5\%, and 10\% of the ERA5 climatological standard deviation.
As evidenced by the quantitative results in Table~\ref{tab:std_abla}, our framework maintains consistent performance across various noise levels, demonstrating remarkable error tolerance. 

\begin{table*}[t]
\centering
\caption{Comparison of latent optimization frequencies (skip=2, 4, 8) in reverse diffusion sampling under sparse observation settings (1\% and 5\%).}
{
\begin{tabular}{c|c|cc|cccccc}
\toprule
\multirow{3}{*}{Ratio}& \multirow{2}{*}{Frequency} & \multirow{2}{*}{MSE} &\multirow{2}{*}{MAE} &\multicolumn{6}{c}{WRMSE} \\
 & &  &  & msl & u10 & u700 & v500 & z500 & t850  \\ \cmidrule{2-10}
& 48h background & 0.0505 &	0.1178 & 98.7265&1.2727 &1.9953	&2.4217 &	89.2752 &	0.9310 	\\ \midrule
\multirow{3}{*}{1\% observation} &skip=2& \cellcolor{blue!10}0.0472 & \cellcolor{blue!10}0.1101 & \cellcolor{blue!10}62.4505 & \cellcolor{blue!10}1.1836 &\cellcolor{blue!10}1.8981 &\cellcolor{blue!10}2.2439 & \cellcolor{blue!10}53.6468 & \cellcolor{blue!10}0.9243   \\ 
&skip=4& 0.0518&0.1162& 70.2005&1.2648& 1.9677&2.33571 &  59.6587& 0.9611  \\ 
&skip=8&\cellcolor{gray!20}0.0549 &\cellcolor{gray!20}0.1205 &\cellcolor{gray!20}74.2434 &\cellcolor{gray!20}1.3258 & \cellcolor{gray!20}2.0143 &\cellcolor{gray!20}2.4029 & \cellcolor{gray!20}62.7748 & \cellcolor{gray!20}0.9840 \\ \midrule
\multirow{3}{*}{5\% observation} &skip=2&\cellcolor{blue!10}0.0309 & \cellcolor{blue!10}0.0851&\cellcolor{blue!10}42.3498 &\cellcolor{blue!10}0.8992 &\cellcolor{blue!10}1.5873 &\cellcolor{blue!10}1.7894 & \cellcolor{blue!10}32.8990 & \cellcolor{blue!10}0.8094   \\ 
&skip=4& 0.0370&0.0939  & 49.5966&0.9892& 1.6697&1.8892 &  41.5379& 0.8538 \\ 
&skip=8&\cellcolor{gray!20}0.0415 &\cellcolor{gray!20}0.1001&\cellcolor{gray!20}54.9766 &\cellcolor{gray!20}1.0740 &\cellcolor{gray!20}1.7544 &\cellcolor{gray!20}2.0039 &\cellcolor{gray!20}45.8147 & \cellcolor{gray!20}0.8890 \\ \bottomrule
\end{tabular}
}
\label{tab:skipabla}
\end{table*}

\textbf{Latent Optimization Frequency Analysis.} 
To empirically validate the theoretical connection (see Section~\ref{sec:laop}) and demonstrate its impact on our framework's effectiveness, we conduct an ablation study on the latent optimization frequency by adjusting the skip interval parameter. 
Table~\ref{tab:skipabla} reveals a systematic performance degradation as the skip interval increases (i.e., fewer optimization steps). 
Specifically, under sparse 5\% observation conditions, the overall MSE increases by 19.74\% and 34.30\% for skip intervals of 4 and 8, respectively, compared to the baseline configuration with skip=2. 
This degradation aligns with our theoretical insight: frequent latent optimization ensures proper integration of observations into the diffusion process, analogous to how iterative refinement in variational DA minimizes the analysis cost function.

\section{Conclusion and Discussion}
\label{sec:conclusion}
\textbf{Conclusion.}
The proposed LO-SDA framework presents a significant advancement in data assimilation by introducing a generative approach that effectively overcomes the limitations of traditional Gaussian assumptions. Building upon the plausible theoretical connection between latent optimization and variational DA methods, we develop a novel framework that integrates observational information into the background conditional prior through latent optimization techniques. Through a combination of latent-conditioned diffusion modeling and such optimization-based observation constraints, LO-SDA provably achieves multiple posterior likelihood optimization during guidance sampling, guaranteeing progressive refinement while demonstrating superior performance compared to both classical variational methods and recent diffusion-based techniques.
Experimental results show that the method achieves substantial improvements in assimilation accuracy, outperforming 3DVAR by 28.25\% and L3DVAR by 14.86\% in z500 WRMSE at 5\% observation coverage. The framework's robustness is further validated under noisy observation conditions, maintaining consistent performance across varying error levels and latent optimization frequencies ablations.

\textbf{Limitations and Impacts. }
The current implementation operates under idealized background conditions, relying on 48-hour forecast backgrounds, which are not compatible with the cyclic DA system. The computational demands of frequent latent optimization also pose challenges for real-time operational use. 
While the architecture naturally extends to 4D DA applications by incorporating model dynamics into the latent optimization process, further research is needed to evaluate its performance under non-Gaussian background error assumptions. 

Despite these limitations, LO-SDA marks a pivotal shift in atmospheric data assimilation, demonstrating that generative models can effectively replace traditional Gaussian assumptions in high-dimensional systems. The framework's success opens new possibilities for nonparametric DA in applications such as climate reanalysis and ensemble forecasting. By merging the flexibility of deep generative models with the rigorous constraints required in operational DA, LO-SDA represents a critical step toward next-generation data assimilation systems. Future work should focus on improving computational efficiency and expanding the framework's applicability to more diverse and realistic atmospheric conditions. 


%

\newpage

\appendix

\section{WRMSE}\label{sec:wrmse}

The latitude-weighted root mean square error (WRMSE) is a statistical metric widely used in geospatial analysis and atmospheric science. Given the estimate $\hat x_{h,w,c}$ and the truth $x_{h,w,c}$, the WRMSE is defined as,
\begin{equation}
\mathrm{WRMSE}(c) = \sqrt{\frac{1}{H \cdot W} \sum_{h,w} H \frac{\cos(\alpha_{h,w})}{\sum_{h'=1}^{H} \cos(\alpha_{h',w})} (x_{h,w,c} - \hat{x}_{h,w,c})^2}\ .
\end{equation}
Here $H$ and $W$ represent the number of grid points in the longitudinal and latitudinal directions,
respectively, and $\alpha_{h,w}$ is the latitude of point $(h, w)$.


\section{The VAE training and results}\label{sec:vae}

\textbf{Model structure and training} We utilize a transformer-based variational autoencoder framework (VAEformer) to effectively reduce the dimensionality of atmospheric data, mapping high-dimensional fields to a compact latent representation ~\cite{han2024cra5}. The architecture incorporates window-based attention mechanisms ~\cite{liu2021swin} to efficiently model atmospheric circulation patterns. Following the "vit\_large" design paradigm, our implementation features identical encoder and decoder structures employing 4×4 patch embeddings with matching stride, a 1024-dimensional latent space, and a 24-layer transformer network utilizing window attention. The model was trained on ERA5 reanalysis data spanning 1979-2016, with the subsequent two-year period (2016-2018) serving as validation, over the course of 60 training epochs.

\textbf{Results} Our trained VAE achieves 0.0067 overall MSE and 0.0486 overall MAE. The varibles WRMSE are presented in Table~\ref{tab:vaermse}.

\begin{table}[htbp]
\centering
\caption{The VAE training results on WRMSE}
\resizebox{\textwidth}{!}{
\begin{tabular}{cccccccccc}
\toprule
u10 & v10 & t2m & msl & z50 & z100 & z150 & z200 & z250 & z300 \\ 
0.54832 & 0.50501 & 0.82944 & 34.002 & 75.529 & 55.645 & 42.436 & 38.426 & 35.963 & 35.008 \\ \midrule
z400 & z500 & z600 & z700 & z850 & z925 & z1000 & q50 & q100 & q150 \\ 
31.623 & 28.4 & 25.948 & 24.563 & 23.415 & 24.37 & 27.266 & 9.64E-09 & 6.35E-08 & 4.90E-07 \\ \midrule
q200 & q250 & q300 & q400 & q500 & q600 & q700 & q850 & q925 & q1000 \\ 
3.04E-06 & 1.02E-05 & 2.47E-05 & 7.81E-05 & 1.68E-04 & 2.73E-04 & 4.01E-04 & 6.02E-04 & 5.95E-04 & 4.69E-04 \\  \midrule
u50 & u100 & u150 & u200 & u250 & u300 & u400 & u500 & u600 & u700 \\ 
0.91052 & 1.1085 & 1.3769 & 1.5108 & 1.5418 & 1.5148 & 1.3712 & 1.2184 & 1.1193 & 1.0552 \\ \midrule
u850 & u925 & u1000 & v50 & v100 & v150 & v200 & v250 & v300 & v400 \\ 
0.95107 & 0.78308 & 0.60413 & 0.80967 & 0.91698 & 1.1081 & 1.2589 & 1.3588 & 1.3544 & 1.2148 \\ \midrule
v500 & v600 & v700 & v850 & v925 & v1000 & t50 & t100 & t150 & t200 \\ 
1.0672 & 0.96791 & 0.90445 & 0.84409 & 0.71597 & 0.55351 & 0.59292 & 0.64698 & 0.47627 & 0.39086 \\ \midrule
t250 & t300 & t400 & t500 & t600 & t700 & t850 & t925 & t1000 & \\ 
0.39115 & 0.41718 & 0.47806 & 0.48918 & 0.50497 & 0.5381 & 0.64627 & 0.61494 & 0.66865 & \\ \bottomrule
\end{tabular}
}
\label{tab:vaermse}
\end{table}

\section{Resampling}\label{sec:resample}
Here we provide a derivation of Equation~\ref{eq:res}. Assume we have two independent Gaussian distributions:
\begin{align}
    p_a(x) &= \mathcal{N}(x; \mu_a, \sigma_a^2) \\
    p_b(x) &= \mathcal{N}(x; \mu_b, \sigma_b^2)
\end{align}
The product distribution $p_c(x) = p_a(x)p_b(x)$ is also Gaussian with parameters:
\begin{align}
    \mu_c &= \frac{\mu_a/\sigma_a^2 + \mu_b/\sigma_b^2}{1/\sigma_a^2 + 1/\sigma_b^2} \\
    \sigma_c^2 &= \frac{1}{1/\sigma_a^2 + 1/\sigma_b^2}
\end{align}

\textit{Proof:} 
\begin{align}
    p_c(x) &\propto \exp\left(-\frac{(x-\mu_a)^2}{2\sigma_a^2}\right) \exp\left(-\frac{(x-\mu_b)^2}{2\sigma_b^2}\right) \nonumber\\
           &= \exp\left(-\frac{1}{2}\left(\frac{1}{\sigma_a^2} + \frac{1}{\sigma_b^2}\right)x^2 + x\left(\frac{\mu_a}{\sigma_a^2} + \frac{\mu_b}{\sigma_b^2}\right) + C\right)
\end{align}
where $C$ contains terms independent of $x$. Completing the square, we obtain:
\begin{align}
    p_c(x) &\propto \exp\left(-\frac{(x-\mu_c)^2}{2\sigma_c^2}\right)
\end{align}
with $\mu_c$ and $\sigma_c^2$ as defined above.

Now consider the conditional distribution in Equation~\ref{eq:res}, where:
\begin{align}
    p(\boldsymbol{z}_t|\hat{\boldsymbol{z}}_0(\boldsymbol{y}),\boldsymbol{y}) &\sim \mathcal{N}\left(\mu(t)\hat{\boldsymbol{z}}_0(\boldsymbol{y}), \sigma^2(t)\boldsymbol{I}\right) \\
    p(\tilde{\boldsymbol{z}}_t|\boldsymbol{z}_t, \hat{\boldsymbol{z}}_0(\boldsymbol{y}), \boldsymbol{y}) &\sim \mathcal{N}(\boldsymbol{z}_t, \lambda_t^2\boldsymbol{I})
\end{align}

The posterior distribution is given by:
\begin{align}
    p(\boldsymbol{z}_t=\boldsymbol{a} | \tilde{\boldsymbol{z}}_t, \hat{\boldsymbol{z}}_0(\boldsymbol{y}), \boldsymbol{y}) 
    &\propto p(\tilde{\boldsymbol{z}}_t | \boldsymbol{z}_t=\boldsymbol{a}) p(\boldsymbol{z}_t=\boldsymbol{a}|\hat{\boldsymbol{z}}_0(\boldsymbol{y}),\boldsymbol{y}) \nonumber\\
    &\propto \exp\left(-\frac{\|\boldsymbol{a}-\tilde{\boldsymbol{z}}_t\|^2}{2\lambda_t^2}\right) 
             \exp\left(-\frac{\|\boldsymbol{a}-\mu(t)\hat{\boldsymbol{z}}_0(\boldsymbol{y})\|^2}{2\sigma^2(t)}\right) \nonumber\\
    &\propto \exp\left(-\frac{1}{2}\left(\frac{1}{\lambda_t^2} + \frac{1}{\sigma^2(t)}\right)\|\boldsymbol{a}\|^2 
             + \left\langle \boldsymbol{a}, \frac{\tilde{\boldsymbol{z}}_t}{\lambda_t^2} + \frac{\mu(t)\hat{\boldsymbol{z}}_0(\boldsymbol{y})}{\sigma^2(t)} \right\rangle \right)
\end{align}

Applying the product formula for Gaussians, we obtain:
\begin{align}
    p(\boldsymbol{z}_t | \tilde{\boldsymbol{z}}_t, \hat{\boldsymbol{z}}_0(\boldsymbol{y}), \boldsymbol{y}) 
    &\sim \mathcal{N} \left( \frac{\lambda_t^2 \mu(t) \hat{\boldsymbol{z}}_0(\boldsymbol{y}) + \sigma^2(t) \tilde{\boldsymbol{z}}_t}{\lambda_t^2 + \sigma^2(t)}, 
    \frac{\lambda_t^2 \sigma^2(t)}{\lambda_t^2 + \sigma^2(t)} \boldsymbol{I} \right)
\end{align}

\section{More visualization.}
We provide additional visualization results comparing different assimilation methods under 5\% observation. In all appendix figures, the top row (left to right) displays the ERA5 ground truth, background field, and background error. The middle row shows assimilation results from (a) our proposed LO-SDA method, (b) the DPS framework, and (c) the Repaint approach, while the bottom row presents the corresponding absolute error fields relative to ERA5 truth. The significantly lighter error magnitudes in the LO-SDA results highlight our method’s superior error reduction capability compared to alternative approaches.

\begin{figure}
    \centering
    \includegraphics[width=1\linewidth]{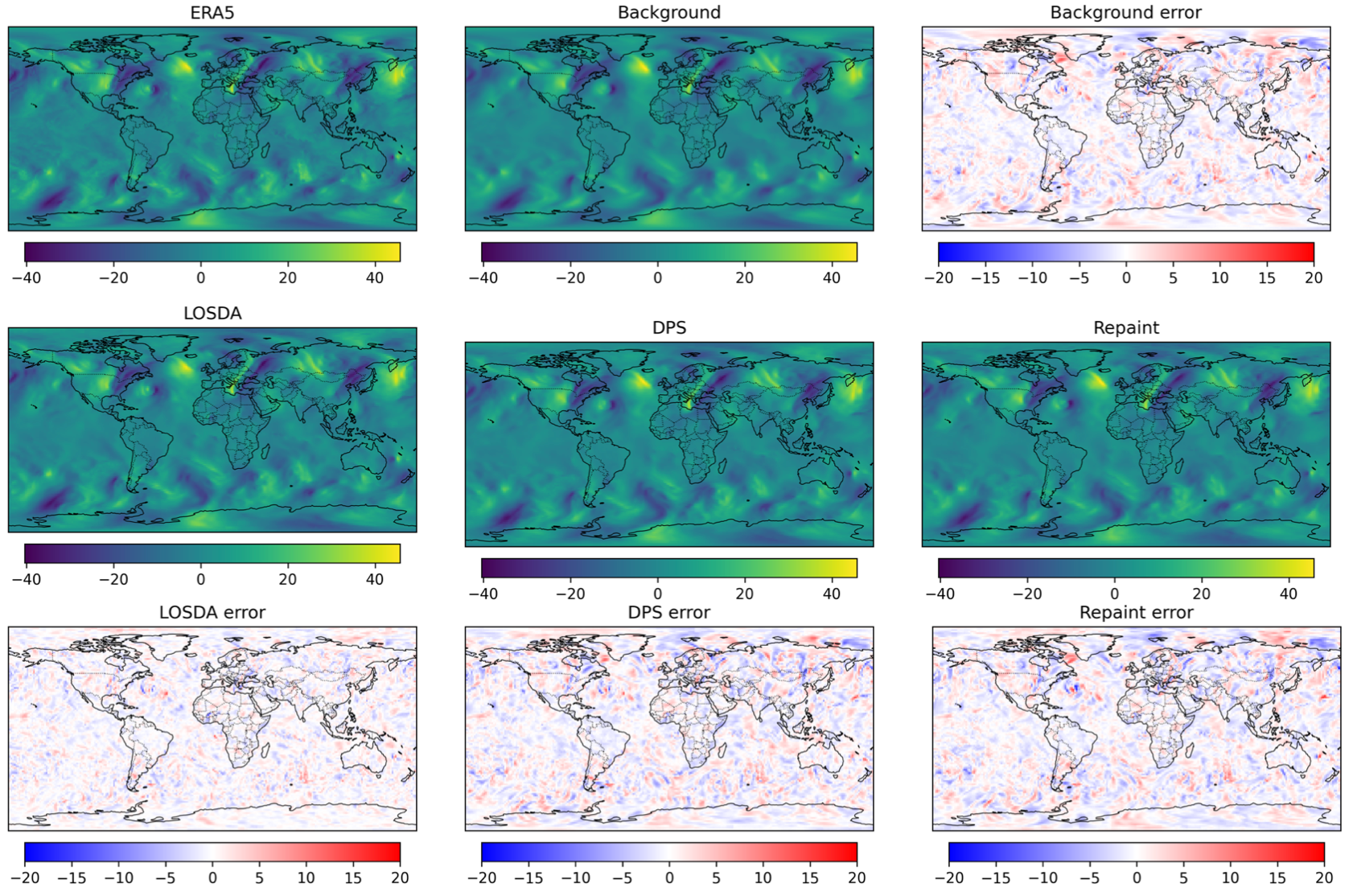}
    \caption{Visulaization of u500 at a 2019-08-26-06:00 UTC.}
    \label{fig:enter-label}
\end{figure}

\begin{figure}
    \centering
    \includegraphics[width=1\linewidth]{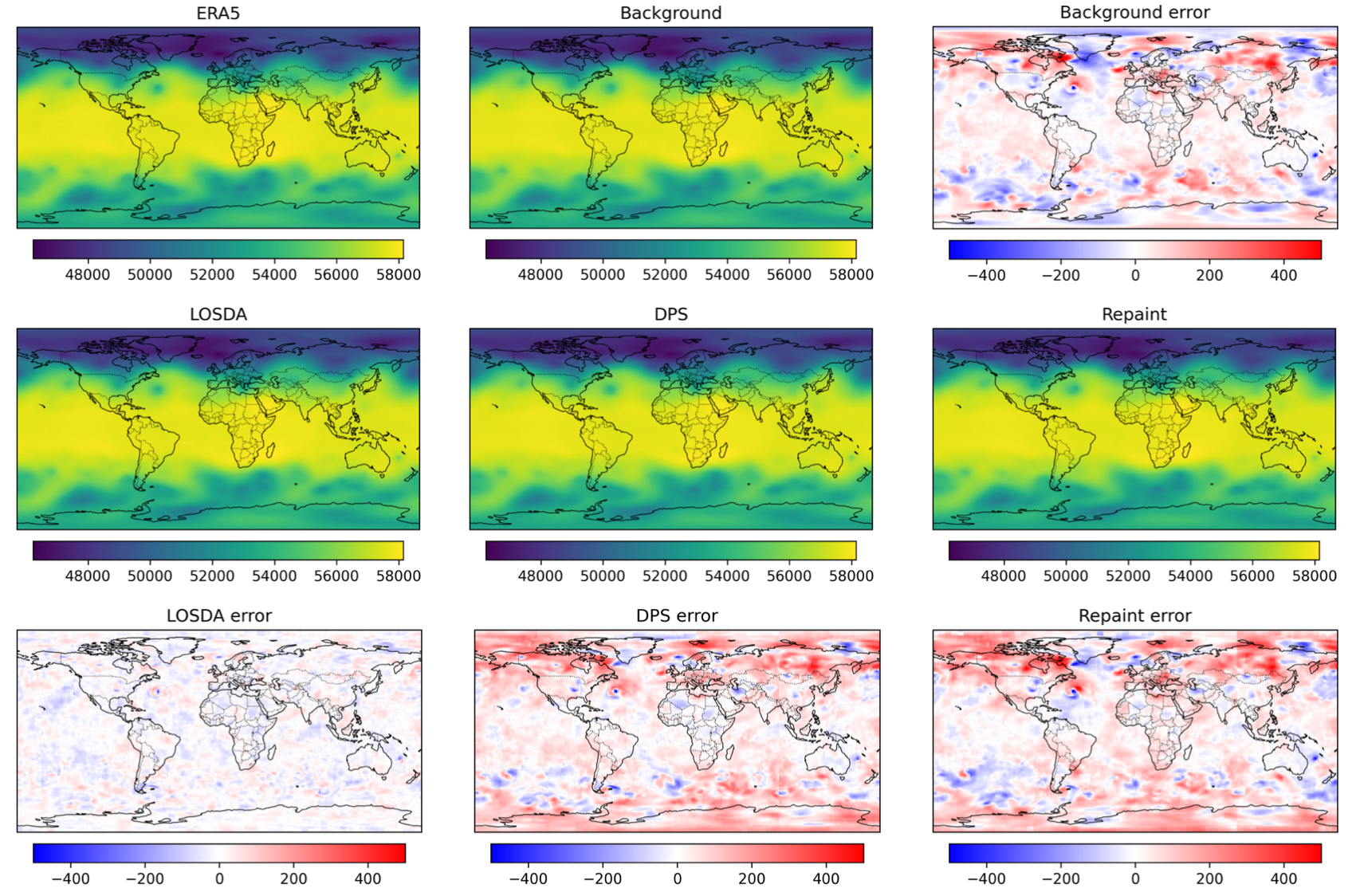}
    \caption{Visulaization of z500 at a 2019-05-18-06:00 UTC.}
    \label{fig:enter-label}
\end{figure}

\begin{figure}
    \centering
    \includegraphics[width=1\linewidth]{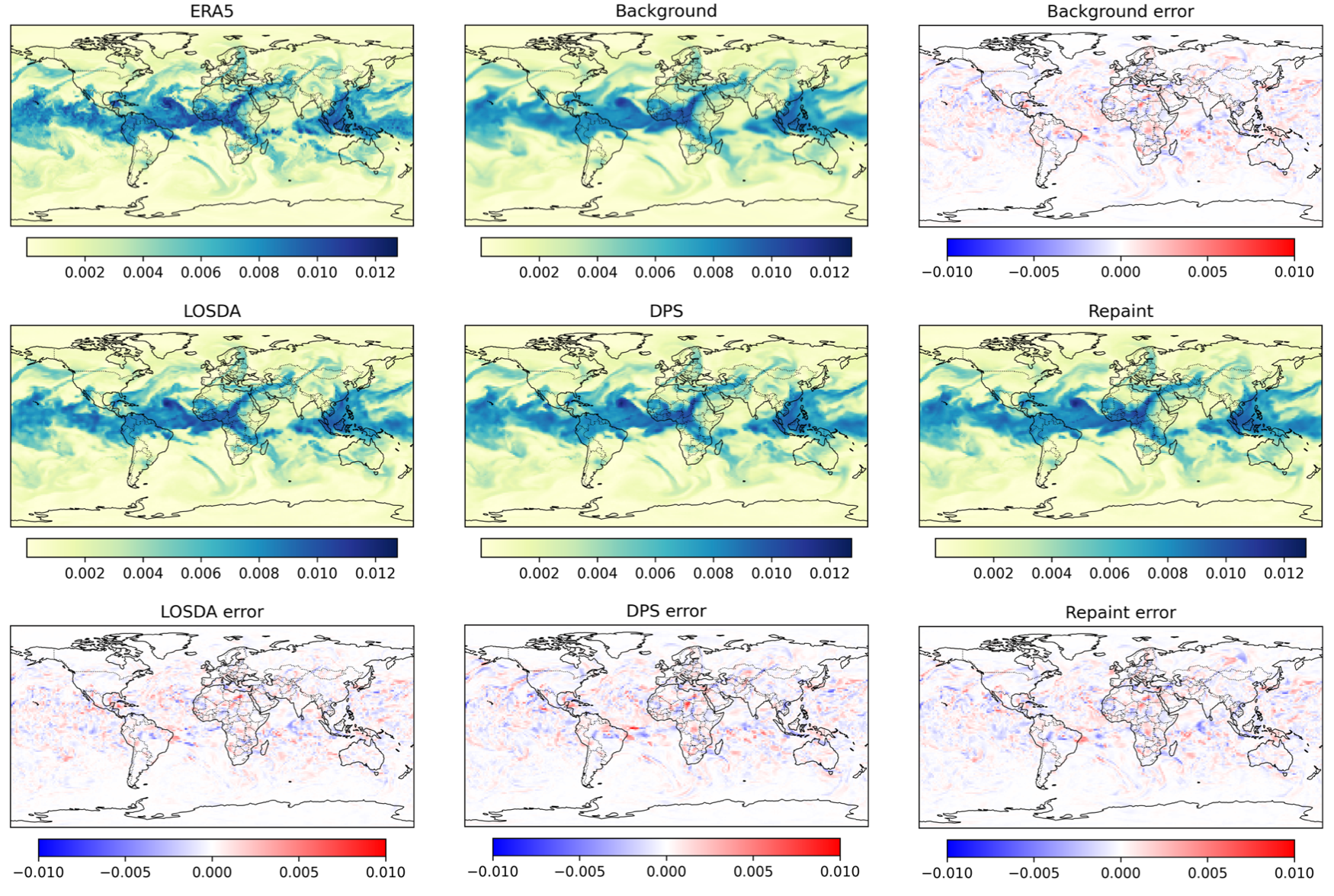}
    \caption{Visulaization of q700 at a 2019-02-02-06:00 UTC.}
    \label{fig:q_vis}
\end{figure}

\begin{figure}
    \centering
    \includegraphics[width=1\linewidth]{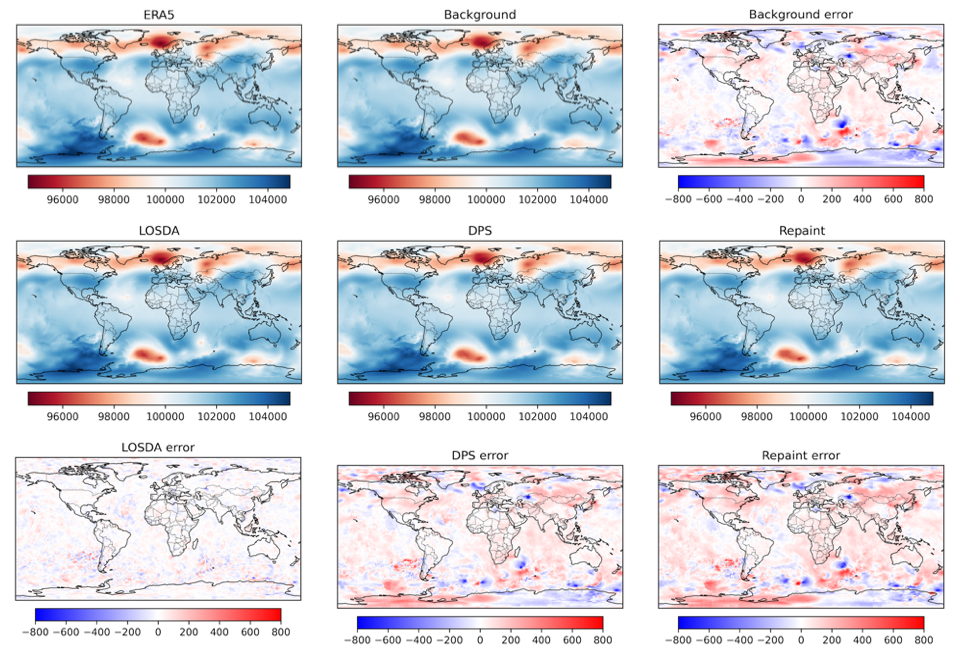}
    \caption{Visulaization of msl at a 2019-04-07-06:00 UTC.}
    \label{fig:msl_vis}
\end{figure}



\newpage
\clearpage

\bibliographystyle{unsrt}
\bibliography{ref}

\end{document}